%% file: arxiv.tex
\documentclass[numbers,sort&compress]{article}
\usepackage{graphicx}
\usepackage{lipsum}
\usepackage{rotating}
\usepackage[normalem]{ulem}
\usepackage{arxiv}
\usepackage{amssymb,amsmath,amsthm}
\usepackage{natbib}
\usepackage{xcolor}
\usepackage{hyperref}
\usepackage[toc,page]{appendix}
\graphicspath{{Artworks/}}
\graphicspath{{imgs/}}
\usepackage{algorithm2e}
\theoremstyle{definition}
\usepackage{url}            %
\usepackage{booktabs}       %
\usepackage{amsfonts}       %
\usepackage{nicefrac}       %
\usepackage{microtype}      %
\usepackage{cleveref}       %
\usepackage{graphicx}
\usepackage{doi}
\input{authors_arxiv.tex}
\date{\today}
\title{Efficient Implementation of Gaussian Process Regression Accelerated Saddle Point Searches with Application to Molecular Reactions}
\hypersetup{
 pdfauthor={},
 pdftitle={Efficient Implementation of Gaussian Process Regression Accelerated Saddle Point Searches with Application to Molecular Reactions},
 pdfkeywords={},
 pdfsubject={},
 pdfcreator={Emacs 30.1 (Org mode 9.7.31)}, 
 pdflang={English}}
\begin{document}

\maketitle
\maketitle
\begin{abstract} %
The task of locating first order saddle points on high-dimensional surfaces describing the variation of energy as a function of atomic coordinates is an essential step for identifying the mechanism and estimating the rate of thermally activated events within the harmonic approximation of transition state theory. When combined directly with electronic structure calculations, the number of energy and atomic force evaluations needed for convergence is a primary issue. Here, we describe an efficient implementation of Gaussian process regression (GPR) acceleration of the minimum mode following method where a dimer is used to estimate the lowest eigenmode of the Hessian. A surrogate energy surface is constructed and updated after each electronic structure calculation. The method is applied to a test set of 500 molecular reactions previously generated by Hermes and coworkers [J. Chem. Theory Comput. 18, 6974 (2022)]. An order of magnitude reduction in the number of electronic structure calculations needed to reach the saddle point configurations is obtained by using the GPR compared to the dimer method. Despite the wide range in stiffness of the molecular degrees of freedom, the calculations are carried out using Cartesian coordinates and are found to require similar number of electronic structure calculations as an elaborate internal coordinate method implemented in the Sella software package. The present implementation of the GPR surrogate model in C++ is efficient enough for the wall time of the saddle point searches to be reduced in 3 out of 4 cases even though the calculations are carried out at a low Hartree-Fock level.
\end{abstract}
\section{Introduction}
\label{sec:introduction}
An important task in the modelling of thermally induced transitions, such as chemical reactions, diffusion events and conformational changes of molecules, is finding the mechanism and estimating the corresponding rate. For transitions that are slow on the time scale of atomic vibrations, the rate can be estimated using transition state theory and when the atoms are in a confined environment vibrating about a well defined average position the harmonic approximation can suffice. Then, the key challenge is finding first-order saddle points on the energy surface describing the variation of the system with respect to the atomic coordinates. At a first-order saddle point, the atomic forces vanish and the Hessian matrix has a single negative eigenvalue. Several methods to systematically identify saddle points have been developed over the past decades, with varying implementations in software packages (see, for example, Refs. \citenum{petersReactionRateTheory2017,henkelmanMethodsFindingSaddle2002,asgeirssonNudgedElasticBand2021,Henkelman2018}). When combined directly with electronic structure calculations, the key issue is to reduce as much as possible the number of times the energy and atomic forces need to be calculated as each calculation may require hours of computations.

There are two types of saddle point searches. Either both the initial state and the final state minima are specified beforehand, i.e. a two endpoint boundary condition, or only the initial state is specified beforehand, i.e. a single endpoint boundary condition. The former is often approached by discretising the path and converging on a minimum energy path connecting the two endpoints, using for example, the nudged elastic band (NEB) method \cite{jonssonNudgedElasticBand1998,asgeirssonExploringPotentialEnergy2018}. The latter is more challenging and can provide unexpected results for the transition mechanism as well as the resulting final state. Such calculations can, therefore be used as the basis for simulations of the long timescale evolution of a system with, e.g. the adaptive kinetic Monte Carlo method \cite{henkelmanLongTimeScale2001}. Furthermore, single endpoint methods can provide an efficient way to complete two endpoint calculations when a rough convergence to the saddle point has been obtained \cite{asgeirssonExploringPotentialEnergy2018}.

Machine learning can be used to accelerate searches for saddle points in many ways. A potential energy surface can be generated to mimic the atomic interactions in a system using various machine learning approaches with input from electronic structure calculations and then the evaluation of energy and atomic forces in saddle point searches is relatively fast. The accuracy of such energy surfaces is, however, often not good enough in regions near saddle points because training sets tend to lack data from these regions. It is then important to keep retraining the energy surface as information on saddle point regions is acquired. The training of a potential energy surface in this manner requires a large set of electronic structure calculations. When many systems are being screened for a particular property, using for example, workflow engines (e.g. AiiDA \cite{huberAutomatedReproducibleWorkflows2022}, Snakemake \cite{molderSustainableDataAnalysis2021}, etc.), the training of a transferable potential energy surface for each candidate is not viable because it requires a large investment in the training set.
It is then better to use machine learning to accelerate each saddle point search using just the minimum number of electronic structure calculations required to reach convergence. This is the approach demonstrated here, namely the acceleration of a single saddle point search so as to reduce the number of electronic structure calculations as much as possible.

A single endpoint calculation is often started near the initial state minimum starting with little or no bias for the possible transition mechanism. A climb up the energy surface is then carried out until a saddle point is reached. Another option is to start from a configuration of the atoms that is likely to be close to a saddle point and thereby attempt to reduce the number of steps needed for the search, at the risk of biasing the calculation towards a preconceived transition mechanism. A database of known saddle point configurations possibly augmented by generative machine learning can be used to then suggest saddle point configurations for systems under study \cite{choiPredictionTransitionState2023,vandevijverKinBotAutomatedStationary2020}. The starting structure may then have significantly higher energy than the saddle point, as will be demonstrated below. In either case, an automated procedure can be used to build a local approximation to the energy surface using input from the electronic structure calculations that need to be carried out as the atomic coordinates are sequentially improved to reach the saddle point geometry. This concept has mostly been used in two endpoint calculations of minimum energy paths, such as the NEB using Gaussian process regression (GPR) \cite{koistinenMinimumEnergyPath2016,koistinenNudgedElasticBand2017} or neural networks \cite{petersonAccelerationSaddlepointSearches2016}. A similar approach can, of course, also be used in local minimization \cite{bisboGlobalOptimizationAtomic2022,denzelGaussianProcessRegression2018a,denzelGaussianProcessRegression2019,denzelHessianMatrixUpdate2020}.

Cartesian coordinates have most often been used in saddle point search algorithms because of the ease of implementation \cite{olsenComparisonMethodsFinding2004,heydenEfficientMethodsFinding2005,kastnerSuperlinearlyConvergingDimer2008,mousseauTravelingPotentialEnergy1998}. However, the various degrees of freedom of a molecule can have a wide range in stiffness, for example, vibration of a strong covalent bond vs. rotation of a methyl group. It can, therefore, be advantageous to use ``internal coordinates'' instead of Cartesian coordinates \cite{schlegelOptimizationEquilibriumGeometries1982}. An automated construction of internal coordinates can be challenging when considering a wide range of atomic structures. This is in particular the case when near linear configurations of three or more atoms are present. In calculations of heterogeneous catalysis involving the reaction of molecules and surfaces of crystals, such linear arrangements of atoms occur frequently. Recently, an elaborate saddle point search method based on nonredundant internal coordinates has been presented and implemented in software called Sella \cite{hermesSellaOpenSourceAutomationFriendly2022}. The algorithm involves a geodesic update  of constrained internal reaction coordinates \cite{hermesGeometryOptimizationSpeedup2021}. When near-linearity in the arrangement of atoms is encountered, ghost atoms are introduced in an automated manner to help define meaningful internal coordinates. The efficacy of the method has been demonstrated in calculations of 500 systems generated using a database approach for predicting saddle point configurations of the atoms. We note that this data set contains only systems where the saddle point searches using Sella converge \cite{hermesSellaOpenSourceAutomationFriendly2022}.

Here, as noted earlier, we present a different approach, namely the use of Cartesian coordinates in combination with a surrogate energy surface generated with GPR. A local approximation to the energy surface is generated using the electronic structure calculations carried out during each saddle point search. The eigenvector corresponding to the lowest eigenvalue of the Hessian, the so-called minimum mode, is used to define the search direction and it is found using the dimer method \cite{henkelmanDimerMethodFinding1999}. We refer to this method as the GPR-dimer. An earlier version of this algorithm has been presented \cite{koistinenMinimumModeSaddle2020}, but the work presented here uses an improved implementation in C++ within the EON software package \cite{chillEONSoftwareLong2014}. We compare the performance of our method to regular dimer calculations without a surrogate surface, as well as the internal coordinate approach of Sella for the 500 system dataset. Our results show that by using the GPR, the number of electronic structure calculations needed to converge to saddle points can be reduced by an order of magnitude and the overall computational time can be reduced in most cases despite the additional overhead even though the electronic structure calculations are carried out at a low Hartree-Fock (HF) level. A performance similar to that of Sella is found even though Cartesian coordinates are used for this dataset of molecular reactions. This is a promising result because our goal is to use GPR-dimer in calculations of molecular reactions on surfaces of crystals, where near linear configurations of atoms are common and the use of internal coordinates would require the introduction of a number of ghost atoms. In some cases, the GPR dimer method converges on saddle points that are closer to the initial guess, both in terms of atomic coordinates and total energy.

The article is organized as follows: In Section \ref{sec:methods}, the saddle point search method and GPR is reviewed for completeness. The performance of saddle point searches for the 500 systems is presented in Section \ref{sec:results}, and in Section \ref{sec:discussion}, the results are discussed. Conclusions are given in Section \ref{sec:conc}.
\section{Methods}
\label{sec:methods}
\subsection{Minimum mode following with a dimer}
\label{sec:org4fc7ab8}
The minimum mode following (MMF) method \cite{henkelmanDimerMethodFinding1999,plasenciagutierrezImprovedMinimumMode2017} is a technique for iteratively moving from some starting configuration of the atoms to a configuration corresponding to a first order saddle point on the energy surface. The search is guided by the eigenvector corresponding to the lowest eigenvalue of the Hessian, i.e. the minimum mode \cite{henkelmanDimerMethodFinding1999}. If at least one of the eigenvalues of the Hessian is negative, the component of the atomic force in the direction of the minimum mode is inverted to form the search direction. Otherwise the search direction is uphill along the minimum mode. The minimum mode can be found in several ways without explicitly constructing the Hessian.

One option is to use a ``dimer'' which consists of two replicas of the system (i.e. configurations of all atoms) separated by a small, fixed distance. Letting the two images be denoted as \(\mathbf{R}_1\) and \(\mathbf{R}_2\), they are positioned on either side of the midpoint \(\mathbf{R}\), such that \(\mathbf{R}_1 = \mathbf{R} + \Delta R \hat{\mathbf{N}}\) and \(\mathbf{R}_2 = \mathbf{R} - \Delta R \hat{\mathbf{N}}\), where \(\hat{\mathbf{N}}\) is a unit vector defining the dimer's orientation and \(\Delta R\) is the distance parameter. Starting from an initial guess for the midpoint \(\mathbf{R}\), for example using a preconceived approximate saddle point configuration, and a randomly oriented unit vector \(\hat{\mathbf{N}}\), the energy of each image, (\(E_1\) and \(E_2\)), with the corresponding atomic forces (\(\mathbf{F}_1\), \(\mathbf{F}_2\)) are evaluated at each image. The dimer is rotated so as to minimize the total dimer energy, \(E\), which is the sum of the energy of the two images, \(E = E_1 + E_2\), while keeping the separation between the images fixed. This aligns the orientation of the dimer, \(\hat{\mathbf{N}}\), with the minimum mode. An iterative process is used to converge to the direction of the minimum mode to a given tolerance in the force \cite{olsenComparisonMethodsFinding2004}. During the rotation phase, the atomic forces are projected to make them perpendicular to the current dimer orientation, \(\mathbf{F}^\perp\). Each rotational step involves finding a plane defined by the rotational force and the previous rotation plane, and the rotational force is then zeroed within this plane. Here, a conjugate gradient (CG) method is used where at each step a linear combination of the force vector and the previous search direction is calculated. The coefficient of the previous direction, or the CG parameter, is also a sensitive indicator of convergence. The Polak-Ribiere method
\cite{polakNoteConvergenceMethodes1969}:

\begin{equation}
\mathbf{G}_i^\perp = \mathbf{F}_i^\perp + \beta_i (\mathbf{G}_{i-1}^\perp + \hat{\mathbf{N}}_{i-1}^{**})
\end{equation}

\begin{equation}
\beta_i = \frac{(\mathbf{F}_i^\perp - \mathbf{F}_{i-1}^\perp) \cdot \mathbf{F}_i^\perp}{\mathbf{F}_i^\perp \cdot \mathbf{F}_i^\perp}
\end{equation}

where \(\hat{\mathbf{N}}_{i-1}^{**}\) refers to a unit vector on the previous plane of rotation, and perpendicular to the current dimer orientation. The number of energy and force calculations is reduced further at the risk of increasing the error in the estimation of the rotational force by using a small angle approximation \cite{heydenEfficientMethodsFinding2005}.

Once the dimer has been oriented sufficiently close to the minimum mode, a translational step moves the midpoint of the dimer forward. If at least one of the eigenvalues of the Hessian is negative, a modified force, \(\mathbf{F}^\dagger\), is used, where the component of the force along the dimer direction is inverted: \(\mathbf{F}_R - 2(\mathbf{F}_R \cdot \hat{\mathbf{N}}) \hat{\mathbf{N}}\). This inversion ensures that the dimer moves uphill on the energy surface along the minimum mode and downhill along all other directions. The dimer is iteratively rotated and translated in this way until the atomic forces become zero to a given tolerance, indicating that a saddle point has been located. The limited memory Broyden-Fletcher-Goldfarb-Shanno algorithm (L-BFGS) on
the midpoint of the dimer at each translation step, as it has previously been found to be efficient \cite{kastnerSuperlinearlyConvergingDimer2008,sheppardOptimizationMethodsFinding2008}. The maximum displacement in each iteration is chosen here to be relatively small, 0.05 \{\AA{}\}, in order to reduce the probability of the dimer search escaping from the saddle point basin of attraction the starting configuration belongs to.
\subsection{Gaussian Process Regression}
\label{sec:org99f5207}
A Gaussian Process Regression (GPR) is a non-parametric method for learning from data. It can be understood as a generalization of multivariate Gaussian distributions to function spaces, effectively providing a distribution over possible functions that fit the observed data. Here, we use a GPR to learn a relationship between Cartesian coordinates of the atoms and their corresponding energy and atomic forces. The GPR model infers this relationship by finding similarities, determined by means of a covariance function (often referred to as the kernel). The expressivity of these models stems from the fact that they may be understood as the infinite limit of a neural network \cite{nealPriorsInfiniteNetworks1996} and that any neural network of a finite width is essentially a GP \cite{yangWideFeedforwardRecurrent2019}.

The inverse-distance modified squared exponential
kernel \cite{koistinenMinimumModeSaddle2020,koistinenNudgedElasticBand2019} is used

\begin{equation}
k_{1/r}(\mathbf{x}, \mathbf{x}') = \sigma_c^2 + \sigma_m^2 \exp\left(-\frac{1}{2} \sum_{i} \sum_{\substack{\ j > i}} \left(\frac{\frac{1}{r_{ij}(\mathbf{x})} - \frac{1}{r_{ij}(\mathbf{x}')}}{l_{\phi(i,j)}}\right)^2 \right)
\end{equation}

where the summations are over all atoms in the system, \(( \mathbf{x}, \mathbf{x}' )\) are the input vectors of atomic coordinates, and \(r_{i,j}(\mathbf{x}) = \sqrt{\sum_{d=1}^3 (x_{i,d} - x_{j,d})^2}\) is the Euclidean distance between atoms \(i\) and \(j\) within configuration \(\mathbf{x}\). \(\phi(i, j)\) is a function specifying the type of an atom pair and \(l_{\phi(i,j)}\) is the corresponding length scale parameter controlling how quickly the covariance changes as a function of the inverse distance between atoms. The parameter \(\sigma_c^2\) is a constant variance term while \(\sigma_m^2\) is a variance scaling factor for the distance-dependent part of the kernel. This form of the kernel is chosen because the inverse distance term,
\(1/r_{ij}\), penalizes configurations where atoms are close to each other. This takes into account that atoms cannot occupy the same point in space, effectively modeling the strong repulsive forces at short distances.

This form of the kernel is chosen because the inverse distance term,
\((\frac{1}{r_{ij}(\mathbf{x})})\), penalizes configurations where atoms
are very close to each other. This incorporates the constraint that atoms cannot occupy the same point in space, effectively modeling the
strong repulsive forces at short distances.

Given this kernel, the GPR prediction of energy and forces for a new configuration takes the form of a Gaussian distribution. The mean of this distribution provides the predicted energy and forces, while the covariance represents the uncertainty in these predictions. The mean function returns a vector containing predicted energy and the \((x,y,z)\) components of the atomic force on each atom.

The GPR-dimer calculation starts out with up to 6 HF calculations as the dimer is rotated to identify the minimum mode direction, but the usual dimer rotation convergence criteria also apply, so for several systems there are fewer HF calculations in this initial phase. These HF calculations constitute the data set for the initial training of the GPR.

The fitting process involves optimization of the kernel hyperparameters. The computations involve an inversion of a matrix of dimensionality \(O(M^{3}N^{3})\) and requires memory on the order of \(O(M^{2}N^{2})\), where \(M\) is the number of electronic structure calculations that are being used as the training set (i.e. ``observations'') and \(N\) is the number of atomic coordinates, three times the number of atoms. The hyperparameters are optimized using the scaled conjugate gradient method \cite{mollerScaledConjugateGradient1993}. A  Cholesky decomposition is used to speed up the prediction of multiple data points \cite{rasmussenGaussianProcessesMachine2006}.
\subsection{Computational specifications}
\label{sec:orgf9f1c47}
The systems analyzed consist of 500 initial configurations of small organic molecules with between 7 and 25 atoms. The electronic structure calculations are carried out at the HF level with a 3-21G basis set using the NWChem \cite{apraNWChemPresentFuture2020} software. The spin unrestricted formalism is used for doublets and spin restricted closed shell formalism is used for the singlet configurations. The self-consistent field (SCF) threshold is set at 10\(^{-8}\) Hartree, and the criterion for convergence to a saddle point is that the norm of the atomic force vector drops below 0.01 eV/\r{A}. The electronic structure calculations carried out here are at the same level and use the same software as those of \cite{hermesSellaOpenSourceAutomationFriendly2022}.

For the GPR-dimer calculations, which are carried out using EON \cite{chillEONSoftwareLong2014}, the geometry verification pre-processing of NWChem is turned off. For both EON and Sella calculations, the ASE \cite{larsenAtomicSimulationEnvironment2017} file I/O interface to NWChem is used, which means no symmetrization or centering takes place within NWChem, unlike the default settings for NWChem. As mentioned above, the conjugate gradient method \cite{heydenEfficientMethodsFinding2005} is used for the dimer rotations and the limited memory Broyden-Fletcher-Goldfarb-Shanno algorithm \cite{nocedalNumericalOptimization2006,kastnerSuperlinearlyConvergingDimer2008} is used for the translation. Since the validity of the surrogate energy surfaces is limited to the region where data is available, early stopping criteria are used to prevent the calculation from venturing too far outside the region \cite{koistinenMinimumModeSaddle2020,koistinenNudgedElasticBand2019}.

In some exceptional cases, 10 out of the 500, saddle point searches are aborted. This occurs when: (1) NWChem reports a segmentation fault, or (2) when 1000 iterations have been taken, or (3) when the energy has increased by 20 eV. This is discussed further in the results section. Most of the aborted searches are due to condition (1) when the surrogate model generates atomic structures for which NWChem reports errors.

The calculations are carried out on the IRHPC \texttt{elja} computer cluster with \texttt{snakemake} as the workflow engine \cite{molderSustainableDataAnalysis2021}. The cluster nodes have Intel Xeon Platinum 8358 (128 @ 2.600 GHz) CPUs which run Rocky Linux release \texttt{8.10}. The Hartree Fock calculations are run with NWChem built with \texttt{spack}, and each run requests 16 cores exclusively for NWChem with \texttt{3 GB} of memory per node. To reduce I/O overheads, the scratch directory was used.  \texttt{dvc} was used to vendor intermediate data, and the final \texttt{csv} data required for the plots along with the \texttt{orgmode} literate programming document \cite{goswamiContinuousIntegrationTeX2021} is part of the associated monorepo.

The NWChem calculations carried out from EON ``cold start'' the HF calculations from generated input files on disk at each step. Sella, however, makes use of the i-Pi \cite{kapilIPI20Universal2019} engine to use previously obtained wave functions as an initial guess and additionally amortize the startup cost associated with running NWChem from scratch on a new set of inputs. Despite this notable disparity in efficiency for the HF calculations, the median increase in time required for the computations using the GPR-dimer is around 9 minutes longer than the Sella calculations.

Similarity measures were computed using the Hausdorff distance as implemented in the iterative rotations and assignment (IRA) shape-matching algorithm \cite{gundeIRAShapeMatching2021}. This was also extended to calculate the root mean square deviation, or RMSD for comparing structures.
\section{Results}
\label{sec:results}
\subsection{Savings by using GPR}
\label{sec:compgprd_id}
\begin{figure}[htbp]
\centering
\includegraphics[width=.9\linewidth]{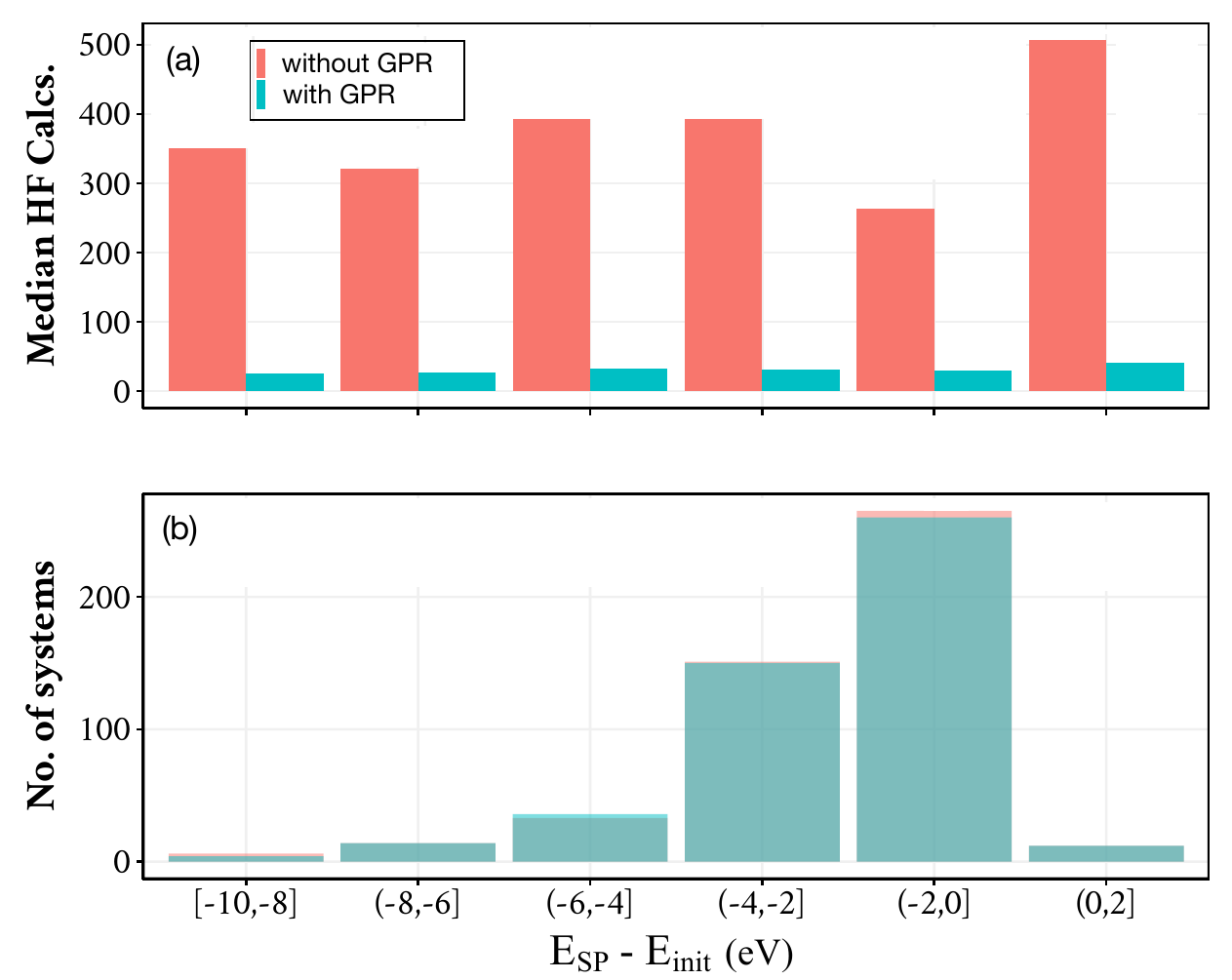}
\caption{\label{fig:gprd_id_hists}(a) Comparison of the median number of HF electronic structure calculations needed to converge on a saddle point, with and without the GPR acceleration, in minimum mode following calculations based on the dimer method applied to the 500 system test set of \cite{hermesSellaOpenSourceAutomationFriendly2022}. The median number of HF calculations drops from 308 to 31 by using the GPR. (b) Number of systems where the difference between the energy of the initial atomic structure and that of the converged saddle point is within each of the 2 eV intervals shown on the horizontal axis. In almost all cases, the saddle point has lower energy than the initial configuration, most often within 2 eV, but there are some examples where the difference is even larger than 10 eV.}
\end{figure}

Figure \ref{fig:gprd_id_hists} shows a comparison of the number of HF calculations needed to reach convergence with and without the use of GPR. The same MMF method with the dimer for finding the minimum mode is used in both cases, but for the GPR-dimer, these calculations are carried out on the surrogate energy surface. An order of magnitude reduction in the number of HF calculations is obtained by using the GPR. The median number of calculations needed for convergence drops from 308 to 31. While HF calculations are relatively fast, this difference in the number of electronic structure calculations translates to large savings in computational effort when higher-level methods are used. The GPR surrogate energy surface construction and overhead are independent of the electronic structure method used for generating the data.

In general, the initial guesses of saddle point configurations of the atoms in the Sella benchmark have higher energy than the saddle points. This is unusual as saddle point searches are more often started near an energy minimum corresponding to an initial state and the saddle point then approached from below. The results in Figure \ref{fig:gprd_id_hists} are arranged in a histogram according to the energy difference between the initial configuration and the obtained saddle point. This energy difference can even be larger than 10 eV for a couple of systems, but these are excluded from the graphic for visual clarity. The number of systems in the data set binned in 2 eV intervals is shown in Figure \ref{fig:gprd_id_hists} (b). Most systems start within 2 eV above the saddle point. Only a few systems start from a configuration that has lower energy than the saddle point. The two sets of calculations, with and without the use of a GPR, most often converge on the same saddle point, but not always, as can be discerned from the color coding of Figure \ref{fig:gprd_id_hists} (b). For example, there are more systems that end up on a saddle point that is within 2 eV of the starting point when the GPR is not used, while there are more systems that end up between 4 and 6 eV below the initial point in the GPR-dimer calculations.

\begin{figure}[htbp]
\centering
\includegraphics[width=.9\linewidth]{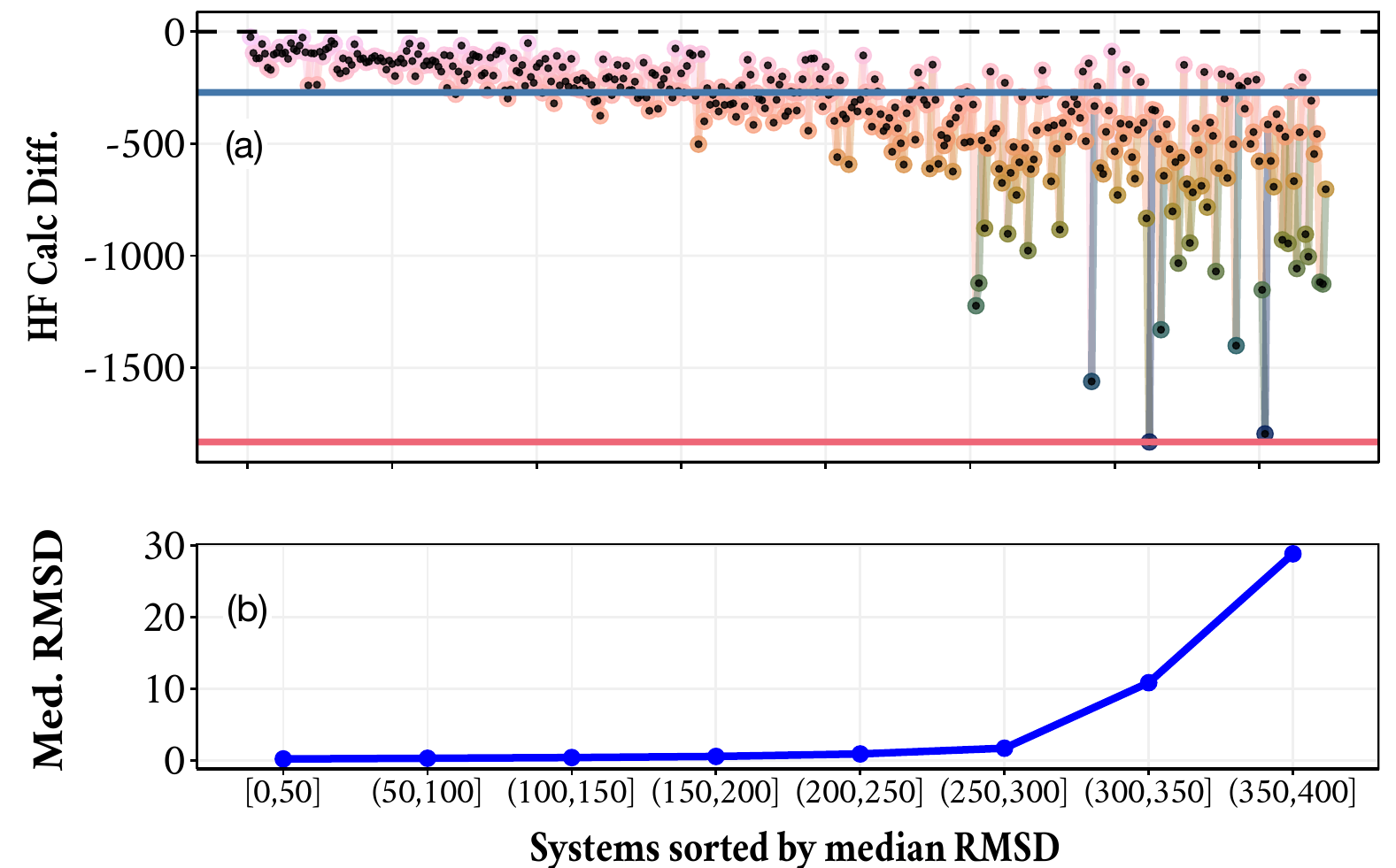}
\caption{\label{fig:gprd_id_pdiff}(a) Difference between the number of HF electronic structure calculations needed to converge on a saddle point with and without the GPR acceleration for each of the systems in the dataset, ordered along the horizontal axis according to the root mean square distance between the initial configuration of the atoms and that of the saddle point found. With the GPR acceleration the median is 29 compared to 301 without it. The mean reduction is indicated by the blue-gray line, 271 HF calculations. Only calculations where the two methods give the same saddle point energy to within 0.01 eV are included, a total of 373 systems. (b) The RMSD distance (in \r{A}) between the initial configuration of the atoms and that of the converged saddle point. Comparison with (a) shows how the efficiency of the GPR acceleration increases with the RMSD distance.}
\end{figure}

More detailed information on this is shown in Figure \ref{fig:gprd_id_pdiff}. There the focus is on systems where the energy difference between saddle points obtained with and without the use of the GPR is less than 0.01 eV, a total of 373 cases. The systems are arranged according to the minimized RMSD of the Cartesian coordinates in the initial and converged configurations of the atoms. The range is from 0 to 30 \r{A}. The savings in the number of HF calculations by using the GPR is clearly greater for larger RMSD values. This is to be expected, since the reduction in the number of HF calculations comes from being able to take several steps on the GPR surface without having to resample from the true energy surface. The GPR-dimer algorithm effectively dynamically adjusts the step size as seen from the perspective of the sampling of the true energy surface. The median number of HF calculations for this set is 29 when using the GPR-dimer while it is 301 without the use of the GPR. The average reduction is 271 HF calculations. However, the distribution is broad, for three cases, where the RMSD is between 10 and 30 \r{A}, the reduction in the number of calculations is over 1500. The savings by using a GPR is particularly large when the initial guess of the saddle point is poor.

\subsection{Comparing GPRD and Sella}
\label{sec:compgprd_sella}
\begin{figure}[htbp]
\centering
\includegraphics[width=.9\linewidth]{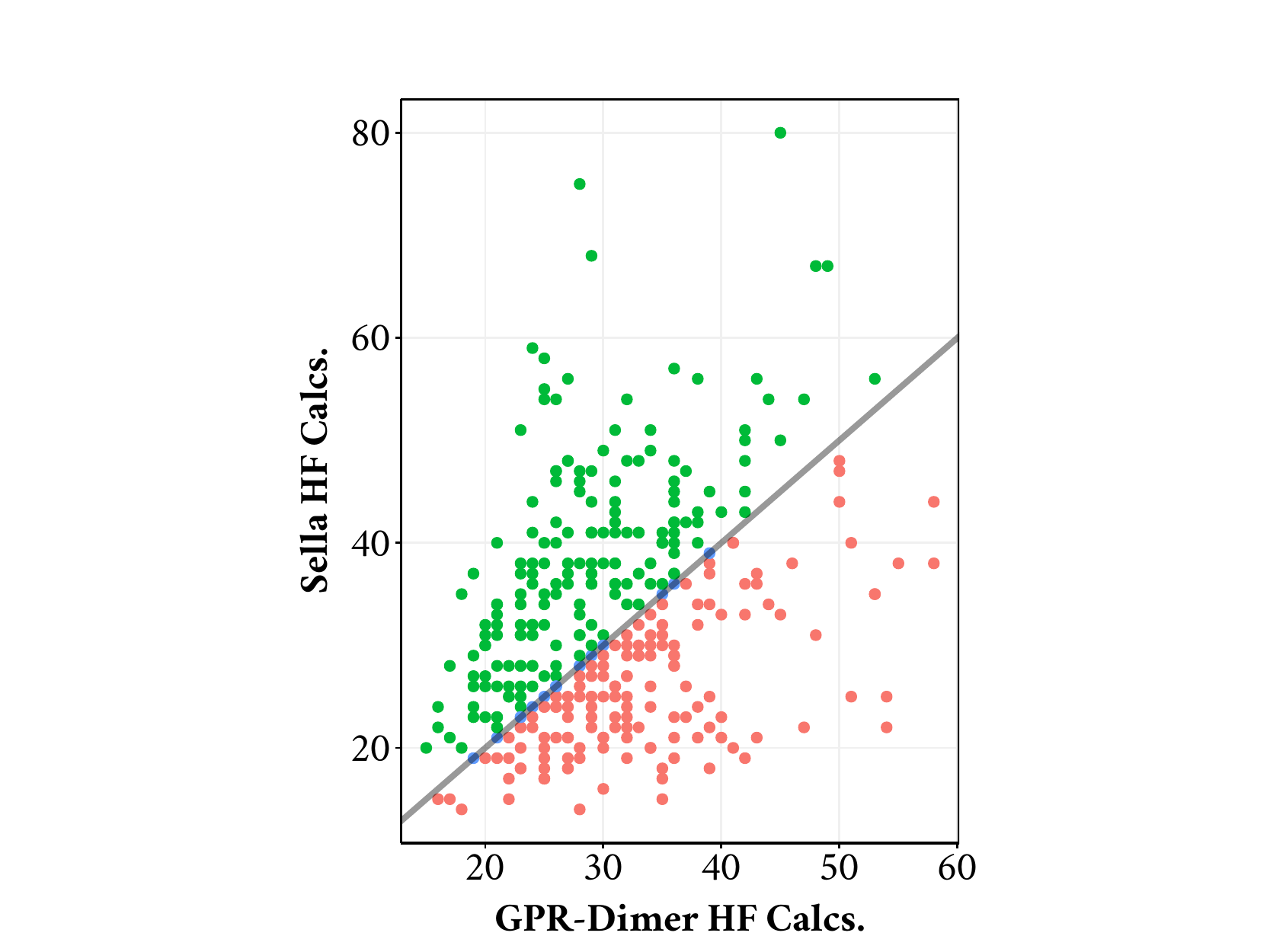}
\caption{\label{fig:gprd_sella_pes}Comparison of the number of HF electronic structure calculations needed to reach convergence using the GPR-dimer method which uses Cartesian coordinates in combination with GPR acceleration, and the Sella method which makes use of internal coordinates. The number of HF calculations is on average similar for both methods, with the GPR-dimer requiring a median of 29 HF calculations while Sella requires 31. The comparison includes 345 systems where the two methods lead to the same saddle point energy to within 0.01 eV. In several cases, the two methods do not converge to the same saddle point, as illustrated in Figure \ref{fig:sella_anom}.}
\end{figure}

Figure \ref{fig:gprd_sella_pes} shows a comparison of the number of HF calculations needed to reach convergence using the GPR-dimer and using Sella. The comparison includes 345 systems where the two methods lead to the same saddle point energy to within 0.01 eV. The number of HF calculations is on average similar for the two methods, GPR-dimer requiring a median of 29 HF calculations while Sella requires 31, despite the fact that the GPR-dimer calculations are carried out using Cartesian coordinates while Sella makes use of internal coordinates which are more natural for the large range in stiffness of the molecular degrees of freedom. The GPR-dimer outperforms Sella in 57\% cases and the mean improvement is 8 HF calculations fewer than Sella.

\begin{figure}[htbp]
\centering
\includegraphics[width=.9\linewidth]{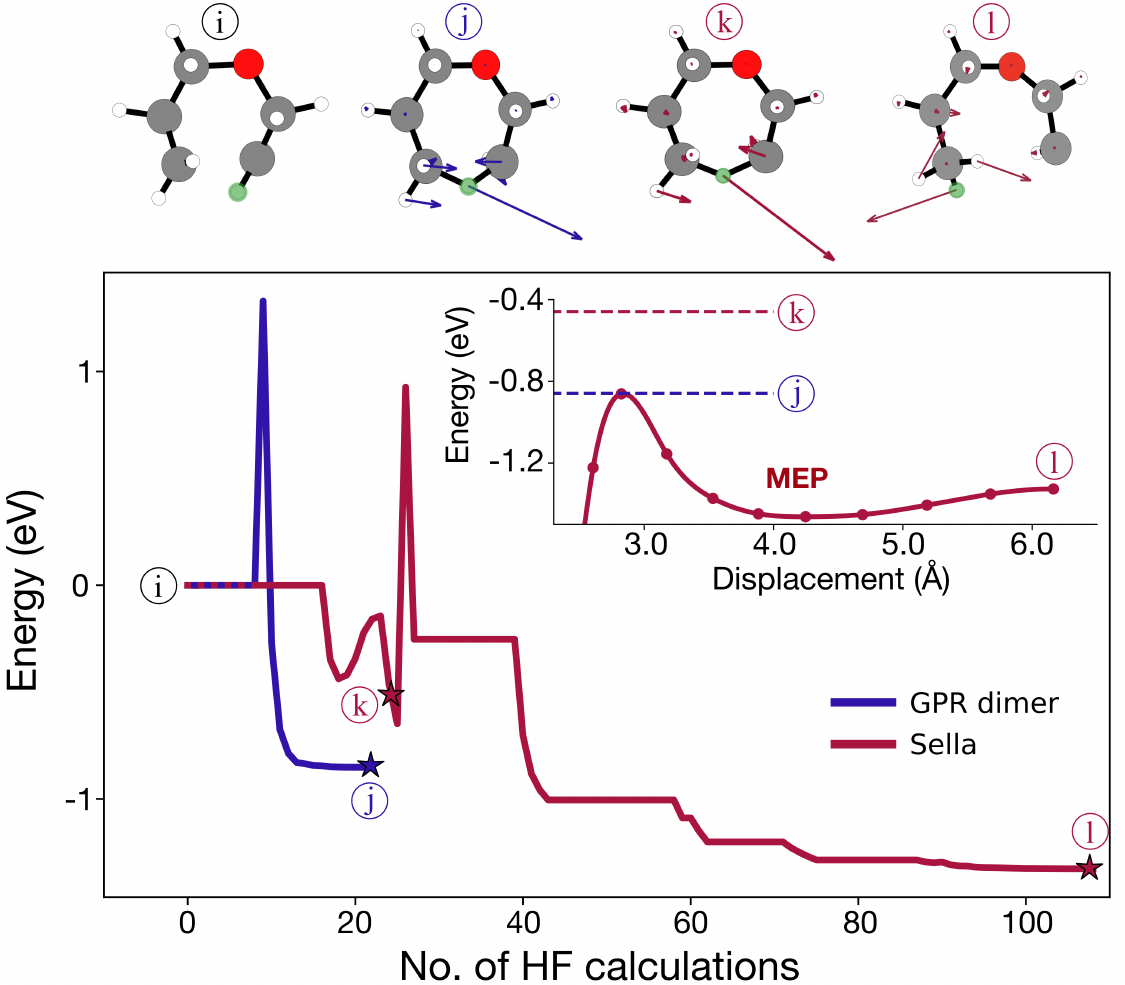}
\caption{\label{fig:sella_anom}Subfigure (a) depicts the path taken during the saddle search for the GP-Dimer and Sella. Subfigure (b) shows the results of an NEB between the initial geometry and the saddle found by each system.}
\end{figure}

There are several cases where the MMF calculation, with or without GPR, does not converge on the same saddle point as the calculation with Sella. This happens, for example, for the first system in the dataset, \texttt{singlet 000}, as illustrated in Figure \ref{fig:sella_anom}. The system in question is a 16-atom molecule with a stoichiometry of \(\mathrm{C}_5\mathrm{OH}_{10}\). The initial configuration of the atoms corresponds to an acyclic ether with a fairly large separation between the carbon endpoints of around 2.2 \r{A}. As can be seen in Figure \ref{fig:sella_anom}, the saddle point found by the GPR-dimer corresponds to the hydrogen from one of the endpoints transitioning to the other. At the saddle point, the C-H bond distances are \(\approx 1.25\) \r{A} and the energy has dropped by 0.9 eV. The RMSD between the initial configuration of the atoms and the saddle point is only 0.2 \r{A}. The Sella calculation comes close to this configuration during the course of the saddle point search, but instead of converging, it proceeds to the final state of the transition and eventually converges on a saddle point for an irrelevant rotation of a methyl group, 0.4 eV lower in energy than the saddle point found by the GPR-dimer method. The RMSD between this saddle point and the initial configuration is 0.6 \r{A}. The GPR-dimer calculation requires 23 HF calculations while the Sella calculation requires 197. This example illustrates how important it is to check the configuration of atoms obtained in a saddle point search. An NEB calculation of the minimum energy path connecting the initial configuration and the saddle point obtained with Sella shows that the saddle point obtained with the GPR-dimer is indeed located along the path and that the final state of the corresponding transition is an intermediate minimum before the methyl group starts rotating in the Sella calculation (see inset in Figure \ref{fig:sella_anom}).

The use of internal coordinates is particularly challenging when a geometry near a linear arrangement of three or more atoms is encountered. There, ghost atoms are introduced algorithmically in Sella to avoid singularities in the internal coordinates. For the 8 systems in the dataset in Figure \ref{fig:gprd_id_pdiff} where this occurs, the GPR-dimer requires 32 HF calculations, close to the average, while Sella requires 44, which is 42\% larger than its average.

The first system within the dataset that involves the introduction of ghost atoms is system \texttt{singlet 016}. It involves two fragments, a methyl group (\(\mathrm{CH}_3\)) and a molecule with two ether linkages (\(\mathrm{C}_4\mathrm{H}_7\mathrm{O}_2\)), i.e.
1-methoxy-2-(2-methoxyethoxy)ethane. The calculation with Sella first increases the distance between the two fragments, but then eventually attaches the methyl group to form a single chain. The saddle point search requires 114 HF calculations and the saddle point reached is 2.4 eV below the initial configuration with an RMSD of 4.2 \r{A}. The GPR-dimer, however, converges after 42 HF calculations to a different saddle point where an H atom jumps onto the methyl group to form
\%\(\mathrm{C}_{4}\)
\(\mathrm{CH}_{4}\)
and a \(\mathrm{CO}_{2}\) dissociates from the molecule. This saddle point is 0.4 eV higher than the one found by Sella and the RMSD with respect to the initial configuration is 1.7 \r{A}. As in the \texttt{singlet 000} case, the path taken by the Sella calculation comes close to the saddle point found by the GPR-dimer but proceeds to a saddle that is lower in energy and farther away from the initial guess for the saddle point configuration.
\section{Discussion}
\label{sec:discussion}
The 500 system data set is constructed in such a way that it only includes cases where Sella converges to some saddle point \cite{hermesSellaOpenSourceAutomationFriendly2022}. For 10 of these systems, the GPR-dimer calculation ends up being aborted using one of the 3 criteria listed in Section \ref{sec:methods}. In most cases, this occurs because the electronic structure calculation fails and the NWChem software reports an error when atoms have come too close together during the saddle point search on the GPR surrogate energy surface. Also, several of the systems in the dataset involve two separate molecular fragments and in a few cases, the saddle point calculation moves them away from each other until the calculations are aborted. The reason for this is that the direction of the minimum mode then corresponds to the uniform displacement of the two fragments with little change in their internal structure. The energy decreases as the fragments move closer together and so the saddle point search moves the fragments in the opposite direction further apart so as to climb uphill on the energy surface. Similar behaviour is observed in the Sella calculations of these systems. However, in most cases where this occurs, the sideways minimization of the energy eventually leads to the search path turning around and the two fragments then eventually come together in a chemical reaction. However, in a few cases, this does not happen and the GPR-dimer calculation is eventually aborted after a maximum number of iterations has been reached. When the initial configuration of the atoms consists of separated molecular fragments, it could make more sense to start with a local minimization of the energy to bring the fragments to an optimal (relative) position, and then start the saddle point search from a configuration of the atoms that is only slightly perturbed from the minimum energy configuration. Here, we have chosen to work with the input as  published by \cite{hermesSellaOpenSourceAutomationFriendly2022}.

The application of a saddle point search method that is
designed to move uphill in energy is actually questionable
when the initial state consists of two molecular fragments
since the uphill climb in energy will likely move the two fragments apart. An NEB calculation of the minimum energy path for such a
case does not run into this kind of a problem, but then the final state of the reaction needs to be specified. Furthermore, the energy surface of the two fragments may be poorly described with a harmonic approximation.
Then, harmonic transition state theory is not expected to provide a good estimate of the rate. In principle, a free energy calculation should be applied in such cases, and, an optimal dividing surface should be determined by maximizing the free energy \cite{johannessonOptimizationHyperplanarTransition2001}.

An important aspect of the use of a surrogate energy surface is not to venture too far from the region of the data points in the training set since extrapolation is not reliable. This is implemented by early stopping criteria in the GPR-dimer saddle point searches \cite{koistinenMinimumModeSaddle2020,koistinenNudgedElasticBand2019}. . The number of times early stopping criteria are applied correlates with longer searches, as the structures added to the training data due to the early stopping are not typically near a saddle point. Here, Euclidean distances have been used to decide on early stopping, but we expect that a Hausdorff distance measure could work better in tandem, and this will be tested in future refinements of the method. Additionally, the Euclidean distance measure is not sensitive to changes across chemical species, which can lead to chemically unfeasible structures being marked as within the cutoff.

The GPR-dimer saddle point search on the surrogate surface can lead to configurations of the atoms that have high energy. The NWChem software used for the HF calculations performs checks on the input structures before calculating the energy and forces, and these are triggered even when the GPR is not used. While the GPR-dimer explores unphysical structures, high energy and atomic forces are in principle not problematic as the GPR can successfully handle such regions. However, NWChem often terminates when atoms come too close and this prevents the GPR-dimer from navigating these regions. We hypothesise that by providing some energy/force value, even a large one representing extreme repulsion, as a pseudo-output from the electronic structure calculation through a wrapper in ASE or EON would allow the GPR-dimer to successfully converge to a saddle point in such cases, while here they represent a failure.

Another improvement that is being pursued for future refinements of the method is the pruning of input data, namely, the elimination of some of the configurations that have been calculated with the electronic structure method early in the search and are far from the saddle point. This will help reduce the computational cost of the matrix inversion and the memory requirement.

In future work we intend to apply these and other methods to larger and more diverse benchmark sets that are not biased towards any particular method. The results of such studies will be reported in an ongoing collaboration to extend the OptBench suite of benchmarks for finding saddle points and minimum energy paths of thermally activated transitions \cite{chillBenchmarksCharacterizationMinima2014}.
\section{Conclusions}
\label{sec:conc}
The results presented here demonstrate that saddle point searches for molecular reactions can be performed efficiently using Cartesian coordinates despite the large disparity in the stiffness of the molecular degrees of freedom if the calculation makes use of GPR to generate a surrogate energy surface. The MMF method is used here in combination with a dimer estimate of the minimum mode. The use of a GPR reduces the number of electronic structure calculations needed to reach convergence by an order of magnitude.

The calculations are carried out for a dataset of 500 systems presented
previously by \cite{hermesSellaOpenSourceAutomationFriendly2022}. It consists of
guesses for saddle points and by construction only includes cases where
calculations using Sella are successful in that some saddle point is obtained. A
comparison is made between the performance of the GPR-dimer method using
Cartesian coordinates and the algorithm in Sella based on internal coordinates
measured in terms of the number of HF calculations required for convergence.

The performance of the two methods is found to be similar.
However, in several cases the Sella calculation bypasses the saddle point that is closest to the initial guess and converges instead on some rearrangement of the atoms in the final state, such as a rotation of a methyl group. This demonstrates the importance of checking the saddle point configuration of atoms obtained in a single endpoint calculation. This can be done efficiently by using the NEB method to find the minimum energy path as it can reveal whether the saddle point found is adjacent to the initial state configuration or whether another saddle point is present in between, as is illustrated in Figure 4.

While the use of a GPR presents a significant overhead for the saddle point search, the efficient implementation presented here in C++ leads to a reduction in the overall computational time even though the electronic structure calculations are carried out here at a low HF level. When higher level electronic structure calculations are used, the overhead will be an even smaller fraction of the total computational effort. The computational effort in the generation of the GPR surrogate energy surface scales rapidly with the dimensionality of the system and the number of electronic structure results included in the training set. Further optimisation is required in order to be able to apply this approach to systems with a large number of movable atoms. Pruning of the input data to drop points far from the saddle point region is one option and the use of only a subset of the atom coordinates of a large system focusing thereby on the region of interest is another, possibly followed by later refinement of the saddle point found with respect to all coordinates.
\section{Acknowledgements}
\label{sec:org368bc2b}
This work was supported by the Icelandic Research Fund (grant no. 217436-053)
and by ``ReaxPro'', funded by the European Union’s Horizon 2020 research and
innovation programme under Grant Agreement No. 814416.

RG thanks Dr. Miha Gunde, Dr. Amrita Goswami, Dr. Moritz Sallermann, Prof.
Debabrata Goswami, Mrs. Sonaly Goswami and Mrs. Ruhila Goswami for discussions.
\section*{Reproduction details}
\label{sec:orgc2bc949}
The authors confirm that the data supporting the findings of this study are available within the article and/or its supplementary materials.
The supplementary material includes: a single repository, which is tagged. These include pinned versions
of ASE, EON (from \url{https://theochemui.github.io/eOn/}), the GPR dimer code, and
a copy of the \texttt{spack} recipes \cite{gamblinSpackPackageManager2015} repository
for building NWChem used for the HartreeFock calculations along with spinout
utilities from Wailord \cite{goswamiWailordParsersReproducibility2022}. A \texttt{conda}
ecosystem based environment managed by \texttt{pixi} is provided to further facilitate
reproducibility, with the modules either pinned to versions or bundled within
the same repository via \texttt{gitsubrepo} following best practices for literate
programming
\cite{goswamiHighThroughputReproducible2022,goswamiReproducibleHighPerformance2022}.

Data related to the results presented in this article and instructions on the generation thereof are available in \cite{goswamiEfficientImplementationGaussian2025MCDat} hosted on the Materials Cloud Archive \cite{talirzMaterialsCloudPlatform2020}.
\section*{Conflict of interest}
\label{sec:orgeab1175}
The authors declare no conflict of interest.
\bibliography{gprd_sella,manual}
\end{document}

%% file: authors_arxiv.tex
\author{ \href{https://orcid.org/0000-0002-2393-8056}{\includegraphics[scale=0.06]{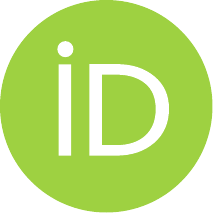}\hspace{1mm}Rohit Goswami}\thanks{Corresponding Author} \\
	Science Instiute\& Faculty of Physical Sciences\\
	Univesity of Iceland\\
	107 Reykjavík, Iceland\\
	\texttt{rgoswami@ieee.org} \\
	\And
	Maxim Masterov \\
	SURF, Science Park\\
	140, 1098 XG Amsterdam, The Netherlands\\
	\And
	Satish Kamath \\
	SURF, Science Park\\
	140, 1098 XG Amsterdam, The Netherlands\\
	\And
	\href{https://orcid.org/0000-0001-8745-2912}{\includegraphics[scale=0.06]{orcid.pdf}\hspace{1mm}Alejandro Peña-Torres} \\
	Science Instiute\& Faculty of Physical Sciences\\
	Univesity of Iceland\\
	107 Reykjavík, Iceland\\
	\texttt{aptorres@hi.is} \\
	\And
	\href{https://orcid.org/0000-0001-8285-5421}{\includegraphics[scale=0.06]{orcid.pdf}\hspace{1mm}Hannes Jónsson}\thanks{Corresponding Author} \\
	Science Instiute\& Faculty of Physical Sciences\\
	Univesity of Iceland\\
	107 Reykjavík, Iceland\\
	\texttt{hj@hi.is}
}